\documentclass[aps,prl,twocolumn,showpacs,amsmath,amssymb,preprintnumbers,10pt,footinbib,superscriptaddress]{revtex4-1}
\usepackage{epsfig}
\usepackage{amsfonts}
\usepackage{amssymb}
\usepackage{enumitem}
\usepackage[normalem]{ulem}
\usepackage{amsthm}
\usepackage{subfig}
\usepackage{bm}
\usepackage{hyperref}
\usepackage{mathrsfs}
\usepackage{bbm}
\usepackage{cancel}
\usepackage[dvipsnames]{xcolor}
\usepackage{accents}
\usepackage{relsize}
\usepackage{float, slashed, graphicx, amssymb, amsmath}
\usepackage{physics}
\usepackage{shuffle}
\usepackage{tikz}
\usetikzlibrary{hobby,calc,shapes, positioning, backgrounds,trees, decorations, decorations.markings, decorations.pathmorphing, arrows, shapes.geometric, snakes}
\usepackage{tikz-feynman}
\tikzfeynmanset{compat=1.1.0}
\tikzfeynmanset{
graviton/.style={circle, draw=green!60, fill=green!5, very thick, minimum size=7mm}
}
\tikzfeynmanset{
codot/.style={/tikz/shape=circle,
/tikz/fill=red,/tikz/minimum size=0.1cm,/tikz/inner sep=1.8pt}
}
\tikzfeynmanset{sb/.style=
{/tikz/shape=circle,
/tikz/fill=black,
 /tikz/minimum size=0.3cm,/tikz/inner sep=1.pt
 } }
\tikzfeynmanset{myblob/.style=
{/tikz/shape=ellipse,
/tikz/fill=red,
 /tikz/minimum width=0.5cm,
 } }
 \tikzfeynmanset{myblobFF/.style=
{/tikz/shape=circle,
/tikz/fill=black,
 /tikz/minimum width=0.25cm,
 } }
 \tikzfeynmanset{myblob2/.style=
{/tikz/shape=rectangle,
/tikz/fill=red,
 /tikz/minimum width=0.1cm,/tikz/inner sep=1.8pt
 } }
  \tikzfeynmanset{myvertex/.style=
{/tikz/shape=circle,
/tikz/fill=black,
 /tikz/minimum width=0.05cm,
 } }
 \tikzfeynmanset{GR/.style={/tikz/shape=ellipse,/tikz/fill={rgb:black,1;white,2},/tikz/minimum size=0.3cm,} }
\tikzset{box/.pic={\filldraw[fill=black]  (0,0) circle (2.5pt);
				   \filldraw [fill=black] (0.5,0) circle (2.5pt);
			       \draw [line width=5pt] (0,0) -- (0.5,0);}}

\tikzset{wiggle/.style={decorate, decoration=snake}}

 \tikzfeynmanset{HV/.style=
{/tikz/shape=circle,
/tikz/fill={rgb:black,1;white,2},
 /tikz/minimum size=0.01cm,/tikz/inner sep=1.8pt
 } }

\usepackage[T1]{fontenc} % if needed

\def\sc#1{\overline{#1}}
\makeatletter
\newcommand \UPlus {\mathop {\operator@font \uplus }\limits }
\makeatother
\makeatletter
\newcommand \Bigcup {\mathop {\operator@font \bigcup }\limits }
\makeatother
  \def\LabelNote#1{}%\smash{\hbox to\phipt{\raise1ex\hbox{\tiny[#1]}\hss}}}
 \def\Label#1{\label{#1}%
  \smash{\hbox to\phipt{\raise1ex\hbox{\tiny[#1]}\hss}}}
  
  \def\mdot{{\cdot}}

\def\veps{\varepsilon}

\def\nn{\nonumber}

\newcommand{\black}{\color{black}}

\def\spa#1.#2{\left\langle#1\,#2\right\rangle}
\def\spb#1.#2{\left[#1\,#2\right]}

\def\be{\begin{equation}}
\def\ee{\end{equation}}
\def\bea{\begin{eqnarray}}
\def\eea{\end{eqnarray}}  
\allowdisplaybreaks

\newcommand{\npre}{\mathcal{N}}

\newcommand{\tF}{\widetilde F}

\begin{document}

\preprint{
	%NBI-PH-22-22
}

\title{Covariant Compton Amplitudes in Gravity with Classical Spin}
\author{N. Emil J. Bjerrum-Bohr}
\email{bjbohr@nbi.dk}
\affiliation{Niels Bohr International Academy, Niels Bohr Institute,\\ Blegdamsvej 17, DK-2100 Copenhagen, Denmark}
\author{Gang Chen}
\email{gang.chen@nbi.dk}
\affiliation{Niels Bohr International Academy, Niels Bohr Institute,\\ Blegdamsvej 17, DK-2100 Copenhagen, Denmark}
\author{Marcos Skowronek}
\email{marcos_skowronek_santos@brown.edu}
\affiliation{Niels Bohr International Academy, Niels Bohr Institute,\\ Blegdamsvej 17, DK-2100 Copenhagen, Denmark}
\affiliation{Department of Physics, Brown University, Providence, RI 02912, USA}
\begin{abstract} 
We develop a novel amplitude bootstrap technique manifestly free of unphysical poles for classically spinning particles interacting with gravitons utilizing only the double-copy and physical factorization limits. Combined with non-factorization polynomial contact contributions from physical data for Kerr black holes, we can address high-spin-order covariant gravitational Compton amplitudes, identifying a pattern for the amplitude that we believe could extend to all orders in spin. Finally, we outline applications and outstanding questions. 
 \end{abstract}  


\maketitle
%\tableofcontents

\section{Introduction}
Computations of scattering amplitudes for classical gravitational physics enjoy remarkable advancements when evaluated in the context of modern quantum field theory programs \cite{Bjerrum-Bohr:2018xdl,Cheung:2018wkq}, inspired by field theoretic undertakings of general relativity ~\cite{Iwasaki:1971vb,Donoghue:1994dn,Bjerrum-Bohr:2002gqz,Holstein:2004dn,Holstein:2008sx,Neill:2013wsa,Bjerrum-Bohr:2013bxa,Damour:2017zjx}. 
With present observations of gravitational waves from black hole mergers, such developments encourage compelling new routes for precision tests of classical physics predictions of Einstein's theory from conservative scattering \cite{Cristofoli:2019neg,Bern:2019nnu,Antonelli:2019ytb,Bern:2019crd,Parra-Martinez:2020dzs,DiVecchia:2020ymx,Damour:2020tta,Kalin:2020fhe,DiVecchia:2021ndb,DiVecchia:2021bdo,Herrmann:2021tct,Bjerrum-Bohr:2021vuf,Bjerrum-Bohr:2021din,Damgaard:2021ipf,Brandhuber:2021eyq,Bjerrum-Bohr:2021wwt,Bjerrum-Bohr:2022ows,Dlapa:2021npj,Bern:2021yeh,Dlapa:2021vgp,Bern:2022jvn,Dlapa:2022lmu,Bern:2021dqo,Jakobsen:2023ndj,Damgaard:2023ttc,DiVecchia:2023frv,Heissenberg:2023uvo,Adamo:2022ooq} and radiative waveform \cite{Kosower:2018adc,Jakobsen:2021smu,Mougiakakos:2021ckm,Jakobsen:2021lvp,Riva:2022fru,Brandhuber:2023hhy,Herderschee:2023fxh,Elkhidir:2023dco,Georgoudis:2023lgf,Caron-Huot:2023vxl}. A captivating spin-off of this program is considering classical physics for spinning black holes, {\it e.g.}, ~\cite{Guevara:2017csg,Arkani-Hamed:2017jhn,Guevara:2018wpp,Chung:2018kqs,Guevara:2019fsj,Arkani-Hamed:2019ymq,Aoude:2020onz,Chung:2020rrz,Guevara:2020xjx,Chen:2021kxt,Kosmopoulos:2021zoq,Chiodaroli:2021eug,Bautista:2021wfy,Cangemi:2022bew,Ochirov:2022nqz}, as well as ~\cite{Damgaard:2019lfh,Bern:2020buy,Mogull:2020sak,Jakobsen:2021zvh,Comberiati:2022ldk,Vines:2017hyw,Vines:2018gqi,Maybee:2019jus,Haddad:2021znf,Jakobsen:2021smu,Liu:2021zxr,Chen:2022clh,Jakobsen:2022fcj,Menezes:2022tcs,FebresCordero:2022jts,Alessio:2022kwv,Bern:2022kto,Aoude:2022thd,Aoude:2022trd,Damgaard:2022jem,Aoude:2023vdk,Bianchi:2023lrg,Bautista:2023szu}.

Exact knowledge of spin dynamics for binary mergers is crucial for interpreting and analyzing current and future precision gravity wave measurements. It supplies a rich phenomenology because of its distinct signal features; thus, a critical theoretical objective is to design a solid computational framework for yielding a proper post-Minkowskian (PM) expansion of spin effects. In this Letter, we formulate a novel amplitude bootstrap technique manifestly free of unphysical poles for classically spinning particles as a first theoretical step on this path. We envision the results as input for generalized unitarity approaches that  will allow us to test Einstein’s theory at high precision. 

As was confirmed in \cite{Bjerrum-Bohr:2023jau}, the covariant Compton amplitude in the theory of Yang-Mills coupled to a classically spinning particle has a fascinatingly simple structure when expressed covariantly in terms of Abelian field strength tensors, entire functions, and physical propagators. It permits covariant expressions for gravitational scattering to quadratic order in spin utilizing a massless factorization bootstrap method embedded in a double copy procedure. A schematic expression for the two-massive-two-massless graviton amplitude established by our work mentioned above is as follows:
\begin{align}\label{amplitude structure}
    M(1,2, 3, 4)&=-{\npre_a(1,2, 3, 4)\,\npre_0(1,2, 4, 3)\over 2(p_1\mdot p_2)}\nn\\
    &+{\npre_{\rm r}(1,2, 3, 4)\over 4(\sc p_4\mdot p_1) (\sc p_4\mdot p_2)} +\npre_{\rm c}(1,2, 3, 4),
\end{align}
where $p_1$ and and $p_2$ are massless momenta and $\sc p_3$ and $\sc p_4$ are massive. The first term comprised the gravitational numerator derived from the factorization behavior in Yang-Mills theory and was the emphasis of ref. \cite{Bjerrum-Bohr:2023jau}. We assemble it from the double copy employing the classical spin gauge numerator $\npre_a(1,2, 3, 4)$ and its scalar spinless counterpart $\npre_0(1,2, 4, 3)$. These numerator expressions stem from the kinematic algebra framework of, e.g., \cite{Brandhuber:2022enp}, and the classical limit corresponds to maintaining only the leading order in a heavy-mass expansion. Our bootstrap technique for this component thus delivers an important starting point for generating a covariant expression for the gravitational Compton amplitude. We will enhance it - by augmenting its validity to higher spin orders. In our analysis of the piece $\npre_a(1,2, 3, 4)\,\npre_0(1,2, 4, 3)$, we worked exclusively with un-flipped spin propagation (to conform with the double-copy procedure), implying neither spin direction nor magnitude altered during the transition. Nevertheless, from cubic order in spin, there is a requirement to evaluate the effect of a spin-flip correction (not na\"ively supplied by the double copy) symbolized by 
 \begin{align}
{\npre_{\rm r}(1,2, 3, 4)\over 4(\sc p_4\mdot p_1) (\sc p_4\mdot p_2)}.
\end{align}
We can also understand this from the fact both Yang-Mills kinematic numerators $\npre_a(1,2, 3, 4),\, $ $ \npre_0(1,2, 4, 3)$ contain massive propagators see, {\it e.g.}, \cite{Brandhuber:2021bsf,Brandhuber:2022enp,Chen:2022nei}.
We will now demonstrate that we can determine the additional contribution directly from the physical factorization behavior on the massive poles of the gravitational amplitude. The contact contribution
\begin{align}
\npre_{\rm c}(1,2, 3, 4),
\end{align} 
on the other hand, corresponds to amplitude segments not associated with a characteristic physical pole. Such terms originate dynamically from the specific Lagrangian description. We will determine such contributions for a Kerr black hole, order by order, utilizing known classical results and determining an empirical pattern. To our surprise, such derived contact terms appear to have a characteristic form following those derived from the factorization bootstrap. It makes us speculate if it is achievable to extrapolate results to all orders in spin.  
\\[5pt]
We organize the presentation this way: First, we will review and establish a suitable computational setup. We will next devise a bootstrap procedure for the new contributions to the amplitude, thereby improving our initial ansatz \cite{Bjerrum-Bohr:2023jau} and demonstrate how this delivers suggestive expressions for the reminder and contact term corrections mandated at higher spin orders. Finally, we will examine this technology's potential and outline applications.

\section{General structure of classical massive spinning Compton amplitude}
We will graphically illustrate massive spinning Compton amplitude scattering processes as follows.
\begin{align}
\label{compton}
& M(1,\ldots, n{-}2,{n{-}1}, {n}) \nn\\
=&\Bigg(\begin{tikzpicture}[baseline={([yshift=-0.8ex]current bounding box.center)}]\tikzstyle{every node}=[font=\small]	\begin{feynman}
    	 \vertex (a) {\(  \sc p_{ n}=mv\)};
    	 \vertex [right=2.1cm of a] (f2)[GR]{$~~$\textbf{S}$~~$};
    	 \vertex [right=2.5cm of f2] (c){$\sc p_{{n-1}}=mv-q$};
    	 \vertex [above=1.3cm of f2] (gm){};
    	 \vertex [left=0.8cm of gm] (g2){$~~~~~~~~~ p_{1}~~~\cdots$};
    	  \vertex [right=0.8cm of gm] (g20){$ p_{n-2}$};
    	  \diagram* {
(a) -- [fermion,thick] (f2) --  [fermion,thick] (c),
    	  (g2)--[photon,ultra thick,rmomentum'](f2),(g20)--[photon,ultra thick,rmomentum](f2)
    	  };
    \end{feynman}  
    \end{tikzpicture}\Bigg)
\end{align}
considering a particle with mass and spin particle and classical ring radius $a=s/m$ interacting with $n-2$ number of gravitons. We have $q\equiv ( p_{1}+\ldots+ p_{n-2})$.
Everywhere in our formalism $p_1,\ldots,p_{n-2}$ describes momenta of gravitons while $ \sc p_{ {n-1}}$ and $ \sc p_{ n}$ represents momenta for the heavy spinning particle. We specify velocity $v$ in the heavy mass limit by levying $v\cdot v=1$ and $v\cdot q=0$ that follows from $ \sc p_{ n}^2 = \sc p_{{n-1}}^2 = (m v-q)^2  =  m^2$.

The starting point for our improved analysis of the massive spinning Compton amplitude is an ansatz for the general covariant form of the Compton amplitude. We will write it schematically in the following way  (using $x_i\equiv a\mdot p_i, x_a^{ij}\equiv a^2 p_i\mdot p_j$): 
\begin{align}
    &M(1,\ldots, n{-}2,{n{-}1}, {n})\nn\\
    &=\sum_{i,j,k} {1\over \prod_r d_r}g^{(i)}(x_1,\cdots, x_{n-2})\,h^{(j)}(x_{a}^{12},\cdots,x_{a}^{n-3\, n-2}) \nn\\
    &  \,U^{(k)}(a,\sc p_{  n},F_1,p_1,\cdots, F_{n-2},p_{n-2}),
\end{align}
where we have factorized the expressions summed in terms of the characteristic functions $g^{(i)}(x_1,\cdots, x_{n-2})$, $h^{(j)}(x_{a}^{12},\cdots,x_{a}^{n-3\, n-2})$, $U^{(k)}(a,\sc p_{  n},F_1,p_1,\cdots, F_{n-2},p_{n-2})$. The functions $U^{(k)}(a,\sc p_{  n},F_1,p_1,\cdots, F_{n-2},p_{n-2})$ are various tensor products encoding the diffeomorphic-invariant interaction of the massless gravitons with heavy massive spinning particles. For example terms in the $U^{(k)}(a,\sc p_{  n},$ $ F_1,p_1,\cdots, F_{n-2},p_{n-2})$ function takes the schematic form 
\begin{align}
        \Big(\prod_r X_r\mdot F_{i_r}\cdots F_{j_r}\mdot Y_r\Big)   \Big(\prod_r \bar X_r\mdot F_{i_r}\cdots F_{j_r}\mdot \bar Y_r\Big),   
\end{align}
where we use $\cdot$ to denote a covariant index contraction and by $X_r, Y_r, \bar X_r, \bar Y_r$, any vector in $\{a, \sc p_n, p_i\}$ and $F^{\mu\nu}_i=p_i^{\mu}\veps_i^{\nu}-\veps_i^{\mu}p_i^{\nu}$. In the $U(a,\sc p_{  n},F_1,p_1,\cdots, F_{n-2},p_{n-2})$ function, we thus have no additional $x_i$ and $x_a^{ij}$ factors (aside from the ones coming from the contractions with $F_i^{\mu\nu}$), and it is finite by power counting at each perturbative order. The physical propagators $d_r$ takes the form
\begin{equation}
	d_r=
	\begin{cases}
	p_{i_1\ldots i_r}^2 &\text{massless},\\
	\sc p_{n i_1\ldots i_r}^2-m^2 &\text{massive},
\end{cases}
\end{equation}
where we define $p_{i_1\ldots i_r}\equiv p_{i_1}+\ldots +p_{i_r}$. The functions $h^{(j)}(x_{a}^{12},\cdots,x_{a}^{n-3\, n-2})$ are simple monomial functions of $x_a^{ij}$, and the set of functions $g^{(i)}(x_1,\ldots, x_{j})$ %with $x_i\equiv a\mdot p_i$ 
denotes generalized a family of entire functions, which we define in the following way.
\begin{align}
    g^{(i)}(x_1,\ldots, x_{j})\equiv \sum_{r_1, \ldots, r_j} c_i^{(r_1,\cdots,r_{j})}G^{r_1,\cdots,r_{j}}(x_1,\ldots, x_{j}).
\end{align}
\black 
In the above expression, $ c_i^{(r_1,\cdots,r_{j})}$ denotes coefficients we will determine from the bootstrap while the basis functions $G^{r_1,\cdots,r_{j}}$ are defined from: 
\begin{equation}\label{eq:entireBasis}
	G^{r_1,\cdots,r_{j}}(x_1,\ldots, x_{j})\equiv(\prod_{i=1}^{j}\mathcal{O}_i^{r_i})G_{j}(x_1; x_2,\ldots, x_{j}),
\end{equation}
with the $\mathcal{O}_i^{r_i}$ operator defined by \footnote{\footnotesize{We note that alternative bases for the $G^{r_1,\cdots,r_{j}}(x_1,\ldots, x_{j})$ functions are possible, but we find this definition particularly convenient.}}
\begin{align}
   \mathcal{O}_i^{r_i}\equiv \begin{cases}
		1, & r_i=0,\\
		x_i^{r_i}, &r_i>0,\\
		\partial_{x_i}^{r_i},& r_i<0,
	\end{cases}
\end{align}
and where the $G_{j}(x_1; x_2,\ldots, x_{j})$ entire functions are studied in our previous work and explicitly defined by
\begin{align}\label{eq:G1def}
G_1(x_1)&\equiv {\sinh(x_1)\over x_1},\nn\\
	G_2(x_{1}, x_{2})&\equiv {1\over x_{2}}\Big({\sinh(x_{12})\over x_{12}}-\cosh(x_{2}) {\sinh(x_1)\over x_1}\Big),
	\end{align}
and recursively at higher orders
\begin{align}\label{G functions recursive}
	G_j(x_{1},x_{2},\cdots,\,x_{j})&\equiv {1\over x_j}\Big(G_{j-1}(x_{1}{+}x_j;x_{2},\cdots,x_{j-1})\nn\\
 &-G_{j-1}(x_{1},x_{2},\cdots,x_{j-1})\cosh(x_j)\Big).
\end{align}
As we explain in \textit{Supplemental material} (see also \cite{Bjerrum-Bohr:2023jau}), our investigations indicate that products of these elementary functions (which we denote as primary entire functions) conveniently form a natural basis for our bootstrap, as they consistently incorporate the classical spin amplitude's exponential behavior and are manifestly free of spurious poles.

\section{Bootstrap four point Compton amplitude}
We will focus on the four-point gravitational Compton amplitude and set up an ansatz for the reminder term $\npre_{\rm r}$ that only contains the massive propagators. We stress that the term $\npre_{\rm r}$ accounts for the spin-flipping effect. The massless factorization is entirely fixed by the numerator product $\npre_a\times\npre_0$ in (\ref{amplitude structure}), and consequently the remainder terms do not contain any massless poles.
Our starting point is the four-point expression:
\begin{align}
&\npre_{\rm r}(a,\sc p_{4},p_1,p_2,F_1, F_2) \, \nn\\
    &=\sum_{i,j,k} \, g^{(i)}(x_1,x_2) \, h^{(j)}(x^{12}_a) U^{(k)}(F_1,F_2,a,\sc p_4,p_1,p_2).
\end{align}
Here, the indices indicate that we are summing over sets of elements of the respective basis for each type of function $g^{(i)}(x_1,x_2)$, $h^{(j)}(x^{12}_a)$ and $U^{(k)}(F_1,F_2,a,\sc p_4,p_1,p_2)$ that includes all independent elements possible to construct.
Additionally, we will dismiss the super index of $x_a^{12}\equiv x_a$, and remark that $h^{(j)}(x^{12}_a)$ can only be a monomial $(x_a)^j$. %

Regarding the tensor structure for the numerator, we find it convenient to order the terms in even and odd (contributions with no Levi-Civita tensor, and with the Levy-Civita tensor) in the spin variable, {\it i.e.}, $U_{\rm even/odd}(1,2) =\sum_{k, \ {\rm even/odd}} U^{(k)}(F_1,F_2,a,\sc p_4,p_1,p_2)$ and it is possible to deduce the solution%
\begin{align}
   U_{\rm even}(1,2)&= \sc p_4\mdot p_2\Big(\sc p_4^2 (a\mdot F_1\mdot F_2\mdot a) (a\mdot F_2\mdot F_1\mdot \sc p_4)\nn\\
   &+a^2( \sc p_4\mdot F_1\mdot F_2\mdot \sc p_4) (a\mdot F_1\mdot F_2\mdot \sc p_4)\Big),\\
   U_{\rm odd}(1,2) &=(\sc p_4\mdot p_2) (a\mdot F_2\mdot F_1\mdot \sc p_4) \nn\\
   &\times \left((a\mdot F_2\mdot \sc p_4) (a\mdot \tF_1\mdot \sc p_4){-}(a\mdot F_1\mdot \sc p_4) (a\mdot \tF_2\mdot \sc p_4)\right).\nn
\end{align}
In the above expressions, we have utilized the identity 
$\epsilon^{\mu\nu\rho\sigma}\epsilon_{\alpha\beta\gamma\delta} = -\delta^{\mu\nu\rho\sigma}_{\alpha\beta\gamma\delta},$ to rewrite expressions with the spin tensor $S^{\mu\nu}=-\epsilon^{\mu\nu\rho\sigma}\sc p_{4\rho}a_\sigma$ in favor of the Hodge dual of the field strength tensor, represented as $\tF^{\mu\nu}\equiv \frac{1}{2}\epsilon^{\mu\nu\rho\sigma}F_{\rho\sigma}$,
\begin{align}
    p_i\cdot S\cdot \varepsilon_i \to a\cdot \tF_i\cdot \sc p_4. 
\end{align}
We also note the following useful identity
\begin{align}\label{eq:tF2F}
    \tF^{\mu\nu}_i \tF^{\rho\sigma}_j&=-F^{\mu\nu}_jF^{\rho\sigma}_i-(\eta^{\mu\rho} (F_j\mdot F_i)^{\nu\sigma}-\eta^{\mu\sigma} (F_j\mdot F_i)^{\nu\rho})\nn\\
    &-(\eta^{\nu\sigma} (F_j\mdot F_i)^{\mu\rho}-\eta^{\nu\rho} (F_j\mdot F_i)^{\mu\sigma})\nn\\
    &+{1\over 2}(\eta^{\mu\rho}\eta^{\nu\sigma} -\eta^{\mu\sigma}\eta^{\nu\rho})\tr(F_i\mdot F_j).
\end{align}
Once the $U^{(k)}(F_1,F_2,a,\sc p_4,p_1,p_2)$ function is specified, we fix the $G^{r_1,\cdots,r_{n-2}}(x_1,\ldots, x_{n-2})$ and $x_a^j$ contributions in the reminder term from the massive factorization behavior.  To accomplish this, we require a suitable covariant expression for the amplitude residue on the cut $\mathop{\mathrm{Res}}_{p_1\cdot \sc p_4=0}$. Here, according to spinor helicity factorization in \cite{Chen:2021kxt}, we find a consistent product of spin exponentials with scalar amplitudes is
\begin{align}\label{eq:cut_corrected}
    &\hskip-0.1cm \mathop{\mathrm{Res}}_{p_1\cdot \sc p_4=0}M(1,2,3, 4)= \npre_{\rm cut}\equiv 
    (\sc p_4\mdot \varepsilon_1)^2(\sc p_{41}\mdot \varepsilon_2)^2 \nn\\
    &\Big[\!\exp\!\Big(\frac{i\varepsilon_1\mdot S\mdot p_1}{\sc p_4\mdot\varepsilon_1} \Big)\!\exp\Big( \frac{i \varepsilon_2\mdot S\mdot p_2}{\sc p_{41}\mdot \veps_2} \Big)\exp\Big({i\,\tr(F_2F_1S)\over 2(\sc p_4\mdot \veps_1)(\sc p_{41}\mdot \veps_2)}\Big)\Big],
\end{align}
where $\tr(F_2F_1S)$ is shorthand for the cyclic index contraction $(F_2^\mu \cdot F_1\cdot S_\mu).$
This expression is manifestly gauge invariant and free of spurious poles in four dimensions on the support of the cut $p_1\mdot\sc p_4 =0$. 

This high spin massive field's factorization behavior includes classical and quantum contributions. Therefore, although we are ultimately interested in just the classical part of the amplitude, which scales as $\mathcal{O}(m^2)$ \cite{Brandhuber:2021eyq}, there is mixing with the leading order quantum contributions in the cut, and this forces us also to collect the $\mathcal{O}(m^1)$ quantum terms in the ansatz for consistency. An example of the quantum part in the $U^{(k)}(F_1,F_2,a,\sc p_4,p_1,p_2)$ function is 
\begin{align}
    U^{\rm quantum}_{{\rm even}}(1,2)&= \sc p_4\mdot F_1\mdot F_2\mdot \sc p_4 a\mdot F_1\mdot F_2\mdot \sc p_4\,,\nn\\
    U^{\rm quantum}_{{\rm odd  }}(1,2)&= \sc p_4\mdot F_1\mdot F_2\mdot \sc p_4 a\mdot \tF_1\mdot F_2\mdot \sc p_4.
\end{align}

With this, the factorization equations on the massive channel take the form:
\begin{align}\label{eq:cut}
  \Big( \!{ \npre_a(1,2, 3, 4)\,\npre_0(1,2, 4, 3)\over 2p_1\mdot p_2}\!+\!{\npre_{\rm r}\over 4\sc p_4\mdot p_1 \sc p_4\mdot p_2}\Big)\Big|_{\rm cut} \!\!\!=\! \Big({\npre_{\rm cut}\over 2\sc p_4\mdot p_1}\Big).
\end{align}
Here, we organize the reminder contribution to the numerators as follows:
\begin{align}\label{eq:NrAnz}
    \npre_{\rm r}&=\Big(g_{\rm even}(x_1,x_2)(x_a)^{j_{1}}U_{\rm even}(1,2)\nn\\
    &+g^{\rm quantum}_{{\rm even}}(x_1,x_2)(x_a)^{j_{2}}U^{\rm quantum}_{{\rm  even}}(1,2)+\cdots \Big)\nn\\
    &+\Big(1\leftrightarrow 2\Big)+\Big(\text{odd order part}\Big).
\end{align}

Here, the $\cdots$ denotes the summation over the finite number of elements in the basis, which can be determined by considering all the possible combinations of tensor products among the field strength tensors, the momenta, and the spin variable that satisfy all the dimensionality, crossing symmetry and mass scaling conditions.

By substituting eq.\eqref{eq:NrAnz} in eq.\eqref{eq:cut}, we obtain the equations for the $g$-functions.  For even orders in the spin expansion, the factorization behavior equation for the same and opposite helicity structures are, respectively: 
\begin{align}
    &g^{\rm quantum}_{\rm even}(x_1,x_2)(x_a)^{j_{2}}-  g^{\rm quantum}_{\rm even}(x_2,x_1)(x_a)^{j_{2}}+(\cdots)\nn\\
    &= x_2 G_1(x_1) G_1(x_2)-x_1G_1(x_1) G_1(x_2),\nn\\
    &g^{\rm quantum}_{\rm even}(x_2,x_1)(x_a)^{j_{2}}-x_1 x_2 g_{\rm even}(x_1,x_2)(x_a)^{j_{1}}+(\cdots)\nn\\
    &=\sinh \left(x_1-x_2\right)-x_1G_1(x_1) G_1(x_2).
\end{align}
Similarly, for odd orders, the constraints take the following form:
\begin{align}
    &g^{\rm quantum}_{{\rm odd}}(x_2,x_1)(x_a)^{j_{4}}- g^{\rm quantum}_{{\rm odd}}(x_1,x_2)(x_a)^{j_{4}}+(\cdots)\nn\\
    &= {i G_1(x_1) \cosh(x_2)}-{i G_1(x_2) \cosh \left(x_1\right)}, \nn\\
    &(g^{\rm quantum}_{{\rm odd}}(x_1,x_2)-g^{\rm quantum}_{{\rm odd}}(x_2,x_1))(x_a)^{j_{4}}\nn\\
    &+2 x_1 x_2 g_{\rm odd}(x_1,x_2)(x_a)^{j_{3}}+(\cdots)\nn\\
    &=-\frac{i \sinh \left(x_1-x_2\right)}{x_{12}}-2 i \cosh \left(x_1-x_2\right)\nn\\
    &+\frac{i x_2 G_1(x_1) \cosh \left(x_2\right)}{ x_{12}}+\frac{3 i x_1 G_1(x_2) \cosh \left(x_1\right)}{x_{12}}.
\end{align}

As seen, \eqref{eq:cut} fixes the form of the $g^{(i)}$ functions and $x_a$ powers completely, {\it i.e.}, without explicitly performing an expansion in the spin variable $a$. The solution for even orders is:
\begin{align}
   g_{\rm even}(x_1,x_2)&
   =(\partial_{x_1}-\partial_{x_2})\Big(G_1(x_1) G_1(x_2)\Big),\nn\\
   g^{\rm quantum}_{{\rm even}}(x_1,x_2)&=-x_1G_1(x_1)G_1(x_2).
\end{align}
The solution for odd part orders is:
\begin{align}
g_{\rm odd}(x_1,x_2)&=
i(\partial_{x_1}-\partial_{x_2})\Big(G_2(x_1,x_2)\Big),\\
 g^{\rm quantum}_{{\rm odd}}(x_1,x_2)&= {i G_1(x_2) \cosh(x_1)}.\nn
\end{align}
The $x_a$ dependence is trivial for this result since the solution requires $j_{i}=0$ everywhere. All other $U^{(k)}$ functions (than those in the above equations) are seen not to contribute, and thus, the final classical part of the contribution is
\begin{align}\label{eq:NRemain}
    \npre_{\rm r}&=g_{\rm even}(x_1,x_2)\Big(U_{\rm even}(1,2) -U_{\rm even}(2,1)\Big) \nn\\
    &
    +g_{\rm odd}(x_1,x_2)\Big(U_{\rm odd}(1,2) +U_{\rm odd}(2,1)\Big), 
\end{align}
where we have employed the interesting identity: $g_{\rm even}(x_2,x_1)=-g_{\rm even}(x_1,x_2), g_{\rm odd}(x_2,x_1)=g_{\rm odd}(x_1,x_2)$. Verifying that the result is consistent with our previous work regarding the predicted scaling behavior is easy. We recall that the degree of a particular object is defined as its inverse scale order in the spin variable, {\it i.e.,} the maximal number of factors of $1/\chi$ when taking $a\to \chi a$  and $\chi \to i \infty$. For example:
\begin{align}\label{eq:deg}   \deg(G_j)=j,\,\deg(\cosh)=0,\,\deg(\partial_{x_i})=1,\,\deg(a)=-1.
\end{align}
With this definition, one can verify that:
\begin{align}
    \deg(\npre_{\rm r})=0,
\end{align}
meaning that the degree of the reminder numerator is zero. As noted in \cite{Bjerrum-Bohr:2023jau}, we impose this condition because it holds for the three-point amplitude and consequently needs to be satisfied at higher points to yield a correct factorization behavior. An interesting observation is that the classical field strength tensor functions $U_{\rm even/odd}(1,2)$ all vanish identically for the same helicity structure and thus do not change the amplitude with $++/--$ graviton polarisations. This feature is gauge-independent and reflects the universal simplicity of the same helicity configuration.

After determining the piece of the amplitude that contributes to the factorization channels, the only freedom left is in the context of potential contact terms for the amplitude. We will discuss such terms in the next section. 

\section{An empirical ansatz for contact term}
Finally, we will turn to the contact part of the amplitude $\npre_c$. It contains no physical pole; we cannot determine it from the factorization behavior as the remainder contribution. We will only focus on the classical part and again start with the ansatz
\begin{align}
   \npre_c=\sum_{j,r,i,k} g_{[2j+r]}^{(i)}(x_1,x_2)\Big( (x_a)^{j} U^{(k)}_{[-r]}(F_1,F_2,a,p_4,p_1,p_2)\Big).
\end{align}
We will assume, backed by our analysis, that the contact term's scaling degree should be zero. This property is necessary for correct factorization properties. Thus we will keep track of the scaling behaviour (\ref{eq:deg}) and introduce an additional subscript index in $[\,\cdot\,]$ on the functions: $g^{(i)}\to g_{[2j+r]}^{(i)}$ and $U^{(k)}\to U^{(k)}_{[-r]}$. We follow the logic that the function $U^{(k)}_{[-r]}$ and the $(x_a)^{j}$ piece should have a degree less than zero to avoid introducing unphysical poles. Thus, for the entire term to have scaling behavior zero, the function $g_{[2j+r]}$ with those constraints must be of scaling degree order $2j+r$. To constrain the ansatz for the contact term even further, we will also impose regular behavior in the momentum scaling limit ($p_i\to\infty$) as for contributions fixed from physical factorization such as \eqref{eq:NRemain}.  This high-energy behavior is natural when considering a minimally coupled \cite{Arkani-Hamed:2017jhn} theory; see later works in \cite{Cangemi:2023bpe,Brandhuber:2023hhl}. As a consequence, the $x_a$ dependence of the contact terms becomes trivial ($j=0$), and there are no explicit factors of $p_1,p_2$ appearing in the $U^{(k)}$ functions. Thus, power counting constraints imply that the only possible $U^{(k)}$ functions that can appear are of degree $-4$ and we arrive at
\begin{align}\label{gm4}
	\npre_c&=\sum_{i,k} \Big(g_{\rm even\, [4]}^{(i)}(x_1,x_2) U^{(k)}_{\rm even\, [-4]}(F_1,F_2,a, p_4)\nn\\
	&+g_{\rm odd\, [4]}^{(i)}(x_1,x_2) U^{(k)}_{\rm odd\, [-4]}(F_1,F_2,a, p_4)\Big),
\end{align}
where $g_{\rm even\, [4]}^{(i)}(x_1,x_2)$ has to be constructed from basis functions of the type 
\begin{align}\label{g even 4}
	\partial_{x_1}\partial_{x_2}G_1(x_1)G_1(x_2),  &&(\partial_{x_1}^2+\partial_{x_2}^2)G_1(x_1)G_1(x_2),
\end{align}
and similarly for $g_{\rm odd\, [4]}^{(i)}(x_1,x_2)$ 
\begin{align}\label{g odd 4}
	\partial_{x_1}\partial_{x_2}G_2(x_1,x_2), &&(\partial_{x_1}^2+\partial_{x_2}^2)G_2(x_1,x_2).
\end{align}

To fix the unknown coefficients in the deduced contact term minimally, we use physical data derived from the gravitational Kerr scattering by requiring consistency with the 2PM amplitude. We emphasize that our result is in complete agreement with, e.g., \cite{Cangemi:2023bpe}, which derives these contact contributions from gauge invariance.

Fitting the $a^4$ spin order in Kerr black hole solution of gravitational wave scattering \cite{Bautista:2022wjf}, we arrive at
\begin{align}
    &\npre_{{\rm even}\ c}=\Big[{(\partial_{x_1}-\partial_{x_2})^2\over 2!}\Big({G_{1}(x_1)}{G_{1}(x_2)}\Big)\Big]\\
    & \times\Big((a\mdot F_1\mdot \sc p_4) (a\mdot F_2\mdot \sc p_4) (a\mdot F_1\mdot F_2\mdot a)-{a^2\over 2}\nn\\
    &\times ((a\mdot F_1 \mdot F_2\mdot \sc p_4) (a\mdot F_2 \mdot F_1\mdot \sc p_4)\!-\! (a\mdot F_1 \mdot F_2\mdot a) (\sc p_4\mdot F_1 \mdot F_2\mdot \sc p_4))\!\Big),\nn
\end{align}
for the even contact term. We note that the opposite helicity amplitude reduces to the well-known exponential expression in the literature for this order.
Similarly, the odd-order contact term is
\begin{align}
   \npre_{{\rm odd}\ c} &=\Big[{i(\partial_{x_1}-\partial_{x_2})^2\over 2!}\Big({G_{2}(x_1,x_2)}\Big)\Big] \\
   &\times\Big( c_1 \big((a\mdot F_1\mdot F_2\mdot a) (a\mdot F_2\mdot \sc p_4) (a\mdot \tF_1\mdot \sc p_4)-(1\leftrightarrow 2)\big)\nn\\
    &+c_2 \big(a^2 (a\mdot F_2\mdot F_1\mdot \sc p_4) (a\mdot \tF_1\mdot F_2\mdot \sc p_4)-(1\leftrightarrow 2)\big)\Big).\nn
\end{align}
A possible set of values of the remaining free parameters $c_1$ and $c_2$ can obtained by imposing spin-shift symmetry at spin order $a^5$ in the 2PM two-body scattering amplitude \cite{Bern:2022kto,Aoude:2022trd} as
%
%\begin{align}
   $c_1=-{3\over 4},  c_2=0$.
%\end{align}
%
Beyond the $a^5$ order, we find no solution that keeps the shift symmetry in our ansatz of the entire functions, see eq. \eqref{gm4}. It indicates that the concept of shift symmetry proposed for the two-body Hamiltonian potentially could break down.

For a physical Kerr black hole, it is necessary to fit the amplitude to the far-zone gravitational scattering, e.g., the Teukolsky solution \cite{Bautista:2022wjf,Bautista:2023szu}. Then, the $a^5$ free parameter should be set to $c_1=-{1\over 2},  c_2=0$ 
and an extra term\\[-25pt]
\begin{align}\label{eq:extraContact}
	&i \left(p_1\mdot \bar p_4-p_2\mdot \bar p_4\right)\left(a\mdot F_2\mdot \tF_1\mdot \bar p_4+a\mdot \tF_1\mdot F_2\mdot \bar p_4\right)\times\nn\\
	&\Bigg(\frac{\left(a\mdot a\right){}^2  (\bar p_4\mdot F_1\mdot F_2\mdot \bar p_4) }{12 m^2}-\frac{11}{60} (a\mdot a)  (a\mdot F_1\mdot F_2\mdot a) \Bigg)
\end{align}
needs to be included. 

It is important to emphasize that the terms in $\npre_{{\rm even}\ c}, \npre_{{\rm odd}\ c}$ appear correctly at higher order in spin (based on current results up to $\mathcal{O}(a^8)$ in \cite{Bautista:2023sdf}). The absence of certain terms compared to the results in \cite{Bautista:2023sdf} aligns with the terms in eq.~\eqref{eq:extraContact}. Notably, these terms share a common factor $z\equiv {\sqrt{-a\mdot a}\over m} (p_1\mdot \bar p_4-p_2\mdot \bar p_4)$ in \cite{Cangemi:2023bpe}. (This extra parameter can be associated with an effective internal structure \cite{Kim:2023drc}.) By extending our ansatz to include monomials of $z$, the basis of the entire function in eq.~\eqref{eq:entireBasis} also works for the z-dependent terms in \cite{Cangemi:2023bpe}. We refer to the second section in the \textit{Supplemental material} for more details. 

\begin{widetext}
To conclude, a final covariant result with the minimal coupling and high energy scaling behavior that fits the classical 2PM gravitational bending angle up to $a^5$ order is 
\begin{align}
     &M=-\Big(\!\frac{\sc p_3\mdot F_1\mdot F_2\mdot \sc p_3}{p_1\mdot \sc p_3}\Big)\Bigg(\frac{w_1\mdot F_1\mdot F_2\mdot w_2}{2(p_1\mdot p_2) (p_1\mdot \sc p_4)}-\frac{p_1\mdot \sc p_4-p_2\mdot \sc p_4}{4(p_1\mdot p_2)(p_1\mdot \sc p_4)}\times
     \Big(i G_2\left(x_1,x_2\right) (a\mdot F_1\mdot F_2\mdot S\mdot p_2)+i G_2(x_1,x_2) (a\mdot F_2\mdot F_1 \mdot S\mdot p_1) \nn\\
     &+i G_1\left(x_{12}\right) \tr\left(F_1\mdot S\mdot F_2\right)+ G_1(x_1) G_1(x_2) \big( (a\mdot F_1\mdot \sc p_4) (a\mdot F_2\mdot p_1)\!-\!(a\mdot F_1\mdot p_2) (a\mdot F_2\mdot \sc p_4) -\!\frac{p_2\mdot \sc p_4\!-\!p_1\mdot \sc p_4}{2}  (a\mdot F_1\mdot F_2\mdot a)\big)\Big)\Bigg)\, \nn\\
      &+\!{\Big((\partial_{x_1}\!-\!\partial_{x_2})G_1(x_1) G_1(x_2)\Big)\over 4(\sc p_4\mdot p_1) (\sc p_4\mdot p_2)} \Big(\sc p_4\mdot p_2(\sc p_4^2 (a\mdot F_1\mdot F_2\mdot a) (a\mdot F_2\mdot F_1\mdot \sc p_4) +a^2 (\sc p_4\mdot F_1\mdot F_2\mdot \sc p_4) (a\mdot F_1\mdot F_2\mdot \sc p_4))\!-\! (1\leftrightarrow 2)\!\Big)\nn\\
     &+\Big({i(\partial_{x_1}\!-\!\partial_{x_2})G_2(x_1,x_2)\over 4(\sc p_4\mdot p_1) (\sc p_4\mdot p_2)}\Big) \Big((\sc p_4\mdot p_2) (a\mdot F_2\mdot F_1\mdot \sc p_4) ((a\mdot F_2\mdot \sc p_4) (a\mdot \tF_1\mdot \sc p_4){-}(a\mdot F_1\mdot \sc p_4) (a\mdot \tF_2\mdot \sc p_4))+(1\leftrightarrow 2)\Big)\nn\\
      &+\!\Big(\!{(\partial_{x_1}\!-\!\partial_{x_2})^2\over 2!}{G_{1}(x_1)}{G_{1}(x_2)}\Big)\Big((a\mdot F_1\mdot \sc p_4) (a\mdot F_2\mdot \sc p_4) (a\mdot  F_1\mdot F_2\mdot a)\!-\!{a^2\over 2} ((a\mdot F_1 \mdot F_2\mdot \sc p_4) (a\mdot F_2 \mdot F_1\mdot \sc p_4) - (a\mdot F_1 \mdot F_2\mdot a) (\sc p_4\mdot F_1 \mdot F_2\mdot \sc p_4))\!\Big)\nn\\
     &\!\!\!+\Big({i(\partial_{x_1}-\partial_{x_2})^2\over 2!}{G_{2}(x_1,x_2)}\Big)\Big(\! -{1\over 2} \big((a\mdot F_1\mdot F_2\mdot a) (a\mdot F_2\mdot \sc p_4) (a\mdot \tF_1\mdot \sc p_4)\!-\!(1\leftrightarrow 2)\big)\Big)
\end{align}
where $w_i^{\mu}\equiv m\,\cosh(x_i) \sc p_4^{\mu} - \, i \,{} \, G_1(x_i) (p_i\cdot S)^{\mu}\,$. 
\end{widetext}

\section{Conclusion}
This presentation focuses on delivering a covariant form of the classical spin amplitude inferred from the factorization behavior on the massive cut involving a basis of entire functions composed of field strength tensor products that manifest diffeomorphism invariance. Derivations follow the schematic expression for the two-massive-two-massless  graviton \black amplitude in eq.~\eqref{amplitude structure}. 
%
%\begin{align}
 %   M(1,2, 3, 4)&=-{\npre_a(1,2, 3, 4)\,\npre_0(1,2, 4, 3)\over 2(p_1\mdot p_2)} \nn\\
 %   & +{\npre_{\rm r}(1,2, 3, 4)\over 4(\sc p_4\mdot p_1) (\sc p_4\mdot p_2)}+\npre_{\rm c}(1,2, 3, 4).
%\end{align}
%
An essential component of the analysis is that the term ${\npre_{\rm r}(1,2, 3,4)\over 4(\sc p_4\mdot p_1) (\sc p_4\mdot p_2)}$ is zero for same helicity structure contributions, has well-behaved momentum scaling limit and can be shown to be constructed from a basis of primary $g^{(i)}$ functions generated from acting with the differential operator $(\partial_{x_1}-\partial_{x_2})$. We stress that our expression is valid to all orders in spin and thus is suitable even for the critical limit $a\to1$, which usually is observed for super heavy black holes. Inspired by this, we have, based on several empirical observations, derived a contact term $\npre_{\rm c}(1,2, 3, 4)$ associated with the four-point Compton amplitude valid to spin order $a^5$, which we showed to agree with the literature. 

A natural extension of our work involves considering the generalized Compton amplitude with three or more gravitons. Our entire function basis is complete for the generalized Compton amplitude, and applying the bootstrap approach to constrain the amplitude by imposing factorization behaviors, gauge-invariance, and ensuring the absence of any unphysical pole is thus intriguing to investigate. In particular, verifying if scaling behavior constraints can provide information on contact terms consistent with classical Kerr black hole scattering is important. The covariant and manifestly gauge-invariant form of the generalized Compton amplitude with three and four gravitons becomes crucial in binary black hole systems. Utilizing methods like unitarity cuts \cite{Damgaard:2021ipf,Brandhuber:2021eyq,Bjerrum-Bohr:2021wwt,Travaglini:2022uwo,Bjerrum-Bohr:2022blt,Bjerrum-Bohr:2022ows} and heavy-mass/velocity cuts \cite{Brandhuber:2021eyq,Bjerrum-Bohr:2021vuf,Bjerrum-Bohr:2021din}, tree amplitudes can be directly employed to compute classical observables such as the one-loop waveform.   

 It is observed that the factorization behavior of classical pieces does not completely determine the quantum part. To fix it, one needs to consider physical constraints in the massive cut at all orders in mass expansion. However, this might be an overkill, noting that we only had to account for quantum terms in the factorization limit as a consistency requirement. It might be possible to improve the bootstrap method so that only (classical) contributions {\it, e.g.,} as considered in heavy-mass or worldline effective theory are required,  \cite{Damgaard:2019lfh,Brandhuber:2021eyq,Goldberger:2005cd,Porto:2007qi,Mogull:2020sak}. See also, \cite{Bern:2023ity}. 

Another question is related to whether a generalized double-copy structure of the ${\npre_{\rm r}(1,2, 3, 4)\over 4(\sc p_4\mdot p_1) (\sc p_4\mdot p_2)}$ contribution might be possible. In its covariant form, all the classical and related quantum $U^{(k)} $ functions have a common factor $(a\mdot \tF_2\mdot F_1\mdot \sc p_4)$, and we use the relation in eq. \eqref{eq:tF2F} to expand the dual field strength tensor.
The fact that we can extract this factor from all terms in the numerator suggests a potential generalized double-copy structure, but whether this is achievable - is a question we leave for now. 

Our bootstrap procedure still has no direct way of determining the contact terms adequately describing Kerr black holes. Following up on this question, we have speculated if scaling constraints in post-Minkowskian bending angle integrals could be enough to fix it, for instance, from requirements of specific cancellation of divergences under momentum scaling of post-Minkowskian amplitudes and observables. Parallel speculation is a recast of the problem in the context of self-force gravity \cite{poisson2011motion,barack2018self}, see also recent progress from amplitude point of view \cite{Barack:2023oqp,Cheung:2023lnj,Kosmopoulos:2023bwc,Adamo:2023cfp}, since an approach with a background Kerr metric and a test particle with classical spin could be an advantageous way to constrain the contact term more simply than a post-Minkowskian approach. Going beyond the minimal solution when fixing the unknown contact coefficients, it is possible to arrive at a solution that does not, {\it e.g.,} vanish in the same-helicity case but leads to the same second-order post-Minkowskian observables. 
Advancing our knowledge of classical spin effects, especially in the context of understanding the contact term, is an essential problem beyond $a^5$ order. We have elucidated that at $a^5$ order, our ansatz has versatile application even in the case where the Teukolsky equation \cite{Bautista:2022wjf,Bautista:2023szu} is used to determine the contact term. 

\section{Acknowledgements}
We thank Y. F. Bautista, Z. Bern, A. Brandhuber, G. Brown, M. Chiodaroli, T. Damour, J. Gowdy, H. Johansson, J. Kim, K. Haddad, Y. Huang, A. Luna, A. Ochirov, F. Teng, G. Travaglini, T. Wang and C. Wen for their comments. The work of N.E.J.B.-B. and G.C. was supported by DFF grant 1026-00077B and in part by the Carlsberg Foundation. G.C. has received funding from the European Union Horizon 2020 research and innovation program under the Marie Sklodowska-Curie grant agreement No. 847523 INTERACTIONS. M.S. was supported in part by the Stefan and Hanna Kobylinski Rozental Foundation.

\bibliographystyle{apsrev4-1}
\bibliography{KinematicAlgebra}

\newpage 
\renewcommand\appendixname{\LARGE {\textbf {~~~~~Supplemental material}}}
{\center \appendixname}

\section{Primary entire functions}\label{App A}
We define the primary entire function as follows: For a given number $r$ of variables $x_i$, one can consider a summation of terms with $r$ product of a hyperbolic function numerator and with homogeneous polynomials in the denominator. If the summation is free of all finite poles, and all the terms have the same degree defined in (eq.~(26) of main text), we call it a primary entire function. For the Kerr black hole Compton amplitude, we are only interested in the cases where the homogeneous polynomials are products of the variables $x_{i_1\ldots i_j} \equiv x_{i_1}+...+x_{i_j}$. 

For the single variable case, the only primary entire function is 
\begin{align}
    G_1(x_1)&={\sinh(x_1)\over x_1}.
\end{align}

For the two variable cases, there are only two primary entire functions of degree two: 
\begin{align}
    G_2(x_1,x_2)&= {1\over x_{2}}\Big({\sinh(x_{{12}})\over{x_{12}}}-\cosh(x_{2})\, {\sinh(x_{1})\over{x_1}}\Big),\nn\\
     G^{(1)}_2(x_1,x_2)&=G_1(x_i)G_1(x_2).
\end{align}
For two variables, there is no primary function at degree three or higher. So, at high order, the entire functions can only be descendant ones $g(x_1,x_2)$ obtained by acting with derivatives on the primary functions, for instance of the type: 
\begin{align}
\partial_{x_1}^{r_1} \partial_{x_2}^{r_2} G(x_1,x_2).
\end{align}
Here, $r_1$ and $r_2$ denote that we act $r_1$ and $r_2$ times with the given derivative operator, respectively. One can also construct the lower order entire functions as
\begin{align}
	x_1^{r_1} x_2^{r_2}G(x_1,x_2), &&r_i>0.
\end{align}
For the Compton amplitude, our function basis in eq. (9) is thus complete.

For the three variable cases, there are only six primary entire functions of degree three in total:
\newline
\begin{align}
    G_3(x_1;x_2, x_3)&= {1\over x_{2} x_{3}}\Big({\sinh(x_{123})\over x_{123}}{-}\cosh(x_{2}) {\sinh(x_{13})\over x_{13}}
 \nn\\
 -&\cosh(x_{3}) 
  ({\sinh(x_{{12}})\over x_{12}}-\cosh(x_{2}) {\sinh(x_{1})\over x_1})\Big),\nn\\
  G_3(x_2;x_1, x_3)&=G_3(x_1;x_2, x_3)|_{x_2\leftrightarrow x_1},\nn\\
  G^{(1)}_3(x_1, x_2, x_3)&=G_1(x_1)G_2(x_2, x_3)\nn\\
  G^{(2)}_3(x_2, x_1, x_3)&=G_1(x_2)G_2(x_1, x_3)\nn\\
  G^{(3)}_3(x_3, x_1, x_2)&=G_1(x_3)G_2(x_1, x_2)\nn\\
  G^{(4)}_3(x_3, x_1, x_2)&=G_1(x_1)G_1(x_2)G_1(x_3).
\end{align}
Similarly, there is no primary entire function for three variables with degree four and higher. It demonstrates that our function basis in eq. (9) of main text is also complete for the generalized Compton amplitude with three gravitons.
\section{Applications of primary and descendant entire functions}
The primary entire functions and their descendants form a functional basis on which we expand the complete (generalized) Compton amplitude. In a recent publication, \cite{Cangemi:2023bpe}, (released after writing this Letter), the authors assemble the four-point Compton amplitude employing another construction.) The entire functions in their Compton amplitude (equation (20) in \cite{Cangemi:2023bpe}) contain an additional variable $z$, defined as
\begin{align}
z\equiv\frac{\sqrt{-a\mdot a}(p_4\cdot p_2-p_4\cdot p_1)}{m}.
\end{align}
Naively, this extra parameter characterizes a Kerr black hole's ``composite particle degree'' beyond the single-particle field theory. Here, we will show that it is possible to expand their expressions into our basis composed of primary and descendant entire functions.
The first one is given by: 
\begin{align}
	&E(x_1,x_2,z)=\frac{2 e^{x_1-x_2} x_1 \frac{\sinh (z)}{z}-e^{x_1-x_2} \cosh (z)+e^{-x_{12}}}{4 x_1^2-z^2}\nn\\
	&+\frac{-2 e^{x_1-x_2} x_2 \frac{\sinh (z)}{z}-e^{x_1-x_2} \cosh (z)+e^{x_{12}}}{4 x_2^2-z^2}.
\end{align}
When expanding in $z$, only $x_1, x_2$ appear in the denominator. Thus, $E(x_1,x_2,z)$ should only be related to $G_1(x_1),G_1(x_2)$ and their descendants. Indeed, one can see that: 
\begin{align}
	E(x_1,x_2,z)&={e^{-x_2}\over 4}\sum_{j=0}^\infty {z^{2j}\over (2j+1)!}\left({1+\partial_{x_1}\over 2}\right)^{2j+1}G_1(x_1)\nn\\
	&+{e^{x_1}\over 4}\sum_{j=0}^\infty {z^{2j}\over (2j+1)!}\left({1-\partial_{x_2}\over 2}\right)^{2j+1}G_1(x_2)
\end{align}
We write the second one of these functions in the following way: 
\begin{align}
    &\tilde{E}(x_1,x_2,z)=e^{x_1-x_2}\times\nn\\
    &\frac{ \left(\left(\left(x_1-x_2\right){}^2-x_{12}^2+z^2\right) \frac{\sinh (z)}{z}-2 \left(x_1-x_2\right) \cosh (z)\right)}{\left(\left(-x_1+x_2-z\right){}^2-x_{12}^2\right) \left(\left(-x_1+x_2+z\right){}^2-x_{12}^2\right)}\nn\\
    &{+}\frac{\left(\left(x_1{-}x_2\right){}^2+x_{12}^2-z^2\right) \frac{\sinh \left(x_{12}\right)}{x_{12}}+2 \left(x_1{-}x_2\right) \cosh \left(x_{12}\right)}{\left(4 x_1^2-z^2\right) \left(4 x_2^2-z^2\right)}.
\end{align}
The related primary entire functions are  $G_2(x_1,x_2), G_{1}(x_1)G_{1}(x_2)$.
Then $\tilde{E}(x_1,x_2,z)$ can be expanded in the $z$ variable as
\begin{align}
    &\tilde{E}(x_1,x_2,z)= \sum_{j=0}^{\infty}\Bigg(\left(z\over 2\right)^{2j}\times\nn\\
    &{(\partial_{x_1}-\partial_{x_2})^{2j+1}\Big(G_2(x_1,x_2)+G_{1}(x_1)G_{1}(x_2)\Big)\over 4(2j+1)!}\Bigg).
\end{align}

Similarly, the third entire function is 
\begin{align}
	&\mathcal{E}(x_1,x_2,z)=\frac{2 e^{-x_{12}} x_1}{\left(x_{12}\right) \left(4 x_1^2-z^2\right){}^2}-\frac{2 e^{x_{12}} x_2}{\left(-x_{12}\right) \left(4 x_2^2-z^2\right){}^2}\nn\\
	&-\frac{e^{x_1-x_2-z}}{2 z \left(\left(x_1-x_2-z\right){}^2-x_{12}^2\right)}+\frac{e^{x_1-x_2-z} \left(x_1-x_2-z\right)}{z \left(\left(x_1-x_2-z\right){}^2-x_{12}^2\right){}^2}\nn\\
	&+\frac{e^{x_1-x_2+z}}{2 z \left(\left(x_1-x_2+z\right){}^2-x_{12}^2\right)}-\frac{e^{x_1-x_2+z} \left(x_1-x_2+z\right)}{z \left(\left(x_1-x_2+z\right){}^2-x_{12}^2\right){}^2},
\end{align}
and can be expanded as 
\begin{align}
	&\mathcal{E}(x_1,x_2,z)=\sum_{j=0}^{\infty}\Bigg(\left(z\over 2\right)^{2j}\times\nn\\
    &{(j+1)(\partial_{x_1}-\partial_{x_2})^{2j+2}\Big(G_2(x_1,x_2)+G_{1}(x_1)G_{1}(x_2)\Big)\over 4(2j+2)!}\Bigg).
\end{align}

The final entire function, which manifests in dissipative effects, is 
\begin{align}
	&\tilde{\mathcal{E}}(x_1,x_2,z)=-\frac{(-z-1) e^{x_1-x_2-z}}{2 z^2 \left(\left(x_1-x_2-z\right){}^2-x_{12}^2\right)}\nn\\
    &-\frac{e^{-x_{12}} z}{x_{12} \left(4 x_1^2-z^2\right){}^2}+\frac{e^{x_{12}} z}{x_{12} \left(4 x_2^2-z^2\right){}^2}\nn\\
    &+\frac{(z-1) e^{x_1-x_2+z}}{2 z^2 \left(\left(x_1-x_2+z\right){}^2-x_{12}^2\right)}-\frac{e^{x_1-x_2-z} \left(x_1-x_2-z\right)}{z \left(\left(x_1-x_2-z\right){}^2-x_{12}^2\right){}^2}\nn\\
    &-\frac{e^{x_1-x_2+z} \left(x_1-x_2+z\right)}{z \left(\left(x_1-x_2+z\right){}^2-x_{12}^2\right){}^2}.
\end{align} 
Just as all the other ones, the basis of primary functions and their descendants also generates this one: 
\begin{align}
	&\tilde{\mathcal{E}}(x_1,x_2,z)=\sum_{j=0}^{\infty}\Bigg(\left(z\over 2\right)^{2j+1}\times\nn\\
    &{(j+1)(\partial_{x_1}-\partial_{x_2})^{2j+3}\Big(G_2(x_1,x_2)+G_{1}(x_1)G_{1}(x_2)\Big)\over 4(2j+3)!}\Bigg).
\end{align}

\section{Useful relations}
We list some relations for the spin tensor product in gauge $\veps_1\mdot p_2=\veps_2\mdot p_1=0$. The fundamental relation is 
\begin{align}
   &a\mdot \veps _1=\nn\\
   &\frac{\veps _1\mdot \sc p_4 \left(p_1\mdot p_2 a\mdot p_1+p_1\mdot \sc p_4 a\mdot p_1-p_1\mdot \sc p_4 a\mdot p_2+i \lambda_1 p_1\mdot S\mdot p_2\right)}{m^2 p_1\mdot p_2+2 \left(p_1\mdot \sc p_4\right){}^2+2 p_1\mdot p_2 p_1\mdot \sc p_4},\nn\\
    &a\mdot \veps _2=\nn\\
    &\frac{\veps _2\mdot \sc p_4 \left(p_1\mdot p_2 a\mdot p_1+p_1\mdot \sc p_4 a\mdot p_1-p_1\mdot \sc p_4 a\mdot p_2+i \lambda_2 p_2\mdot S\mdot p_1\right)}{m^2 p_1\mdot p_2+2 \left(p_1\mdot \sc p_4\right){}^2+2 p_1\mdot p_2 p_1\mdot \sc p_4}.
\end{align}
It is deduced from the on-shell condition of the polarization vectors $\veps_1\mdot\veps_1=\veps_2\mdot\veps_2=0$. Here, $\lambda_i=\pm1$ denote the positive and negative helicity graviton, respectively. On the massive cut  $p_1\mdot \sc p_4=0$, the relation is further simplified and induces the relations, 
\begin{align}
   \veps_1\mdot S\mdot \veps_2 &\doteq \frac{i (\lambda_1+\lambda_2) a\mdot p_1 \veps _1\mdot \sc p_4 \veps _2\mdot \sc p_4}{m^2},\nn\\
   \veps _1\mdot S_2\mdot \veps _2&\doteq\frac{i \left(\lambda_1+\lambda_2\right) a\mdot p_2 \veps _1\mdot \sc p_4 \veps _2\mdot \sc p_4}{m^2},\nn\\
     \veps _1\mdot S_1\mdot \veps _2&\doteq -\frac{i \left(\lambda_1+\lambda_2\right) a\mdot p_1 \veps _1\mdot \sc p_4 \veps _2\mdot \sc p_4}{m^2},\nn\\
     p_1\mdot S\mdot \veps _1&\doteq-i \lambda_1 a\mdot p_1 \veps _1\mdot \sc p_4,\nn\\
     p_2\mdot S\mdot \veps _2&\doteq -\frac{\veps _2\mdot \sc p_4 \left(p_1\mdot S\mdot p_2+i \lambda_2 \left(m^2 a\mdot p_2+p_1\mdot p_2 a\mdot p_1\right)\right)}{m^2},\nn\\
     p_2\mdot S_1\mdot \veps _2&\doteq\frac{\veps _2\mdot \sc p_4 \left(p_2\mdot S_1\mdot \sc p_4+i \lambda_2 p_1\mdot p_2 a\mdot p_1\right)}{m^2},\nn\\
     p_1\mdot S_2\mdot \veps _1&\doteq\frac{\veps _1\mdot \sc p_4 \left(p_1\mdot S_2\mdot \sc p_4+i \lambda_1 p_1\mdot p_2 a\mdot p_1\right)}{m^2},\nn\\
     \veps _1\mdot \veps _2&\doteq\frac{\left(\lambda_1 \lambda_2+1\right) \veps _1\cdot \sc p_4 \veps _2\cdot \sc p_4}{m^2},
\end{align}
where $S^{\mu\nu}_i=-\epsilon^{\mu\nu\rho\sigma}p_{i\rho}a_\sigma$.
After using all the relations, the amplitude on the cut is only dependent on quantities such as $\veps_i\mdot \sc p_4,\ a\mdot p_i,\ p_1\mdot S\mdot p_2,\ p_1\mdot p_2, a\mdot a$.

Another practical relation follows from the Levi-Civita tensor. We find 
\begin{align}
    X\mdot \tF_1 \mdot  \tF_2 \mdot Y= X\mdot F_2 \mdot  F_1 \mdot Y-{1\over 2} X\mdot Y \tr(F_1\mdot F_2).
\end{align}
This relation can be used to transform terms such as $f(F_k) (a\mdot F_i\mdot F_j\mdot \sc p_4),\ f(F_k) (a\mdot \tF_i\mdot F_j\mdot \sc p_4),$ $f(F_k) (a\mdot F_i\mdot \tF_j\mdot \sc p_4),\ f(F_k) (a\mdot \tF_i\mdot \tF_j\mdot \sc p_4)$ to  $f'(F_k) (a\mdot \tF_2\mdot F_1\mdot \sc p_4)$ using $a\mdot \sc p_4=0$.

\end{document}